# Evaluating the Effectiveness of Mobile Game-Based Learning for Raising Adolescent Health Awareness: The Case of "AHlam Na 2.0"

Noel Caliston

Iloilo State University of Fisheries Science and Technology
Iloilo, Philippines
✉ npcaliston@gmail.com*

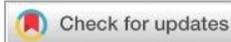

**Abstract**

This study addresses a critical gap in adolescent health education strategies in the Philippines, as highlighted by the Young Adult Fertility and Sexuality (YAFS) survey series, which overlooks the use of games as a medium for disseminating health information. To bridge this gap, the research introduces AHlam Na, a game-based mobile application designed to enhance adolescents' awareness and understanding of key health-related topics. Using a single-group pretest-posttest design, the study involved forty junior high school students from a randomly selected school in the Philippines. They interacted with the application that embedded adolescent health topics into its gameplay. Data collected through pretest and post-test surveys revealed a significant improvement in the student's knowledge and attitudes toward adolescent health after engaging in the game, indicating that game-based learning effectively enhances their learning experience. The positive reception and knowledge gains suggest that AHlam Na is a promising tool for promoting adolescent health awareness. Based on these findings, it is recommended that the application be integrated into the adolescent health curriculum in schools across the Philippines. Future studies should examine the long-term impact of game-based learning on health behaviors and expand the sample size to include more diverse demographic groups. This research contributes to the growing body of literature on game-based learning in health education by demonstrating the potential of digital games to address the limitations of traditional teaching methods. The successful implementation of AHlam Na underscores the importance of exploring gamified learning tools to deliver critical health information to young people effectively.

## A. Introduction

Adolescent sexual initiation significantly increases the risk of teenage pregnancy, which poses considerable risks to both the mother and child (Lara & Abdo, 2016). These risks often result in compromised health, educational setbacks, and diminished future employment prospects (Hindin et al., 2016). According to 2017 United Nations estimates, the Philippines is one of the Southeast Asian countries with adolescent fertility rates exceeding the regional average, with 44 births per 1,000 women aged 15-19 between 2015 and 2020 (Habito et al., 2019). Although recent years have seen a decline in teenage pregnancy and early marriage rates in Western Visayas (Region 6), the Philippines, the issue remains a significant concern. In 2020, the Commission on Population and Development (POPCOM) reported that 8,409 adolescent females between the ages of 10 and 19 became pregnant and gave birth. This trend is particularly alarming, as adolescent parents are often ill-prepared for the challenges of parenthood, perpetuating a cycle of poverty and violence that can span generations.

The Young Adult Fertility and Sexuality (YAFS) survey series serves as a vital resource, providing comprehensive demographic data on adolescent health in the Philippines. It offers valuable insights into Filipino adolescents' and young adults' knowledge and behaviors related to sexuality, reproductive health, education, employment, health, and lifestyle choices. According to YAFS4 and YAFS5, the primary sources of information on sexual health for Filipino adolescents include social media, educational materials, movies, TV, radio, and printed media. However, mobile games are notably absent from the survey as a source of information on sexual health, despite the fact that most adolescents in the Philippines own cell phones, which they use for both study and leisure (University of the Philippines Population Institute, 2016).

Mobile game-based learning is increasingly recognized as an innovative, low-cost, low-risk, and engaging educational approach. Research suggests that games can offer a more enjoyable and effective learning experience than traditional classroom instruction, largely because players control the pace of their learning, whereas, in schools, it is often the teacher who dictates the pace (Vega et al., 2022). This emerging field of study is gaining traction among educators and researchers for its potential to make learning more accessible and enjoyable (Rincón et al., 2022). Recognizing this potential, the Iloilo State College of Fisheries, in collaboration with the Commission on Population and Development Region 6, pioneered the development of the first adolescent health game application in the Philippines. The original version, named "AHlam Na," was designed to deliver essential sexual health education to young Filipinos. The app's name plays on the Filipino phrase "alam na," meaning "already know," symbolizing the aim to empower adolescents with knowledge (Commission on Population and Development & Iloilo State College of Fisheries, 2022).

While the initial release of AHlam Na received varied feedback from stakeholders, no formal study was conducted to evaluate its effectiveness. This feedback, however, laid the groundwork for the development of AHlam Na 2.0, an enhanced version that addresses the shortcomings of the first iteration. This paper evaluates the effectiveness of AHlam Na 2.0 in delivering adolescent health education and aims to fill the gap in research on mobile game-based learning in the context of sexual health education for Filipino youth.

Game-based learning (GBL) involves applying gaming principles in educational contexts to enhance student engagement and motivation. Unlike traditional forms of learning, GBL integrates elements of motivational psychology, leveraging games' fun and interactive nature to make learning more engaging and immersive (Pho & Dinscore, 2015). In the context of adolescent sexual health education, an innovative pedagogy is essential to address the unique needs of a generation immersed in digital technology (Haruna et al., 2018). Digital games, with their ability to captivate attention and deliver content in an enjoyable and memorable way, have increasingly become a popular medium for promoting adolescent sexual health. Recent years have seen a surge in the development and use of such games, focusing on topics like HIV/AIDS prevention, sexually transmitted infections (STIs), vaccinations, and promoting healthy relationship behaviors (Chu et al., 2015; Fiellin et al., 2017; Jiang et al., 2017). These digital interventions have shown significant promise in transforming health education. They provide an efficient, effective, and enjoyable approach to learning, as recognized by scholars (Shegog et al., 2021; Vega et al., 2022) and the "New Media Consortium," which identified digital games as a major trend in the future of e-learning (Giessen, 2015). In particular, many health-related games aim to reduce coercion and peer pressure in adolescent relationships while promoting informed decision-making regarding sexual health (Hieftje et al., 2016; Cates et al., 2018).

Despite the growing adoption of game-based learning for sexual health education, there is still limited research on the effectiveness of these interventions. This study seeks to address this gap by evaluating the impact of a specific game-based approach on adolescent health awareness, contributing to the broader understanding of the role games can play in educational and health-related outcomes.

A valid evaluation of the effectiveness of game-based learning follows a structured approach to ensure reliable results (Anyi, 2019). According to best practices, an experimental design is recommended (All et al., 2016) to assess the impact of the game-based learning intervention. Pre-tests and post-tests were administered and carefully designed to measure learning objectives, ensuring consistency between the two assessments. The pre-test allowed for the evaluation of the participants' baseline knowledge, while the post-test provided data on any learning gains after the intervention. By controlling these variables, the study aimed to provide a clearer understanding of the game's role in enhancing learning outcomes, strengthening the validity of the conclusions drawn.

Understanding user acceptance of mobile digital games is crucial for educators and researchers aiming to integrate these technologies effectively into learning environments (Ghani, 2020). Numerous theoretical models have been introduced to explain how users adopt and decide to use specific technologies. Among these, some of the most widely referenced models include the Theory of Planned Behaviour (TPB) (Alzahrani et al., 2017; Procter et al., 2019), the Technology Acceptance Model (TAM) (Estriegana et al., 2019; Vanduhe et al., 2020), the Innovation Diffusion Theory (IDT) (Rahardja et al., 2019), and the Unified Theory of Acceptance and Use of Technology (UTAUT) (Ramírez-Correa et al., 2019). The TAM is one of the most popular and widely applied frameworks in the study of technology adoption. Originally developed by Davis in 1989, the TAM builds upon the Theory of Reasoned Action, incorporating key elements of this earlier model to explain why users accept or reject particular information technologies. At its core, TAM posits that two primary factors determine technology acceptance: perceived usefulness (PU) and perceived ease of use (PEOU). PU refers to the degree to which a user believes that using a specific technology will enhance their performance, while PEOU reflects the extent to which the user perceives the technology to be free of effort. These factors influence the user's attitudes toward the technology, which in turn affects their behavioral intentions and actual usage.

In the context of this study, the Technology Acceptance Model is particularly relevant as it provides a structured way to evaluate how students perceive and engage with mobile digital games for educational purposes. Ghani (2020) developed a TAM-based questionnaire adapted to assess learning satisfaction in this study. This questionnaire allowed for a comprehensive evaluation of how perceived usefulness, ease of use, and overall satisfaction contribute to the successful adoption of mobile game-based learning. By grounding the study in the TAM framework, the research aims to shed light on key factors that drive user acceptance, ultimately guiding future educational interventions involving digital games.

### B. Research Methods

*Research Participants*

The study recruited a total of 40 high school students, aged between 12 to 17 years, to participate. These students were selected randomly from different academic strands to ensure a diverse representation. The participants spanned across four grade levels, with nine students from the seventh grade, three from the eighth grade, fourteen from the ninth grade, and fourteen from the tenth grade. The detailed distribution of participants by grade level is shown in Table 1.

**Table 1**. Distribution of participants by grade level.

| Grade Level | Frequency | Percentage |
|---|---|---|
| 7 | 9 | 22.5 |
| 8 | 3 | 7.5 |
| 9 | 14 | 35 |
| 10 | 14 | 35 |
| Total | 40 | 100 |

*Procedure*

The experiment began with a pretest administered to the students, lasting 40 minutes. This pretest consisted of fifty items directly related to the content presented in the game. Following the pretest, the students were instructed to download the *"AHlam Na 2.0"* game from Google Play and were given one week to engage with it, with the goal of completing as many levels as possible during that time.

After the one-week gameplay period, the students took a post-test, which contained the same fifty items as the pretest but presented in a different order to avoid memorization effects. Additionally, a learning satisfaction questionnaire was distributed to gauge the participants' overall experience with the game, including their engagement, enjoyment, and perceived learning outcomes.

*Design of the Adolescent Health Mobile Game "AHlam Na 2.0"*

The mobile game utilized in this study, titled *"AHlam Na 2.0,"* was developed as a collaborative project between the Iloilo State College of Fisheries (ISCOF) and the Commission on Population and Development (POPCOM) in Region VI, where the researcher played a major role during the development. This game was designed to serve as an educational intervention aimed at raising awareness about adolescent health topics, including sexual health, healthy relationships, and responsible decision-making.

To ensure the game met the needs of its intended audience, high school students from various parts of the region were invited to participate in the planning and design process. These students provided valuable input regarding the game's features and interface, which helped create a user-centered design that would appeal to adolescents. The development also involved consultations with key government agencies such as the Department of Health (DOH) and POPCOM Region VI, ensuring that the content adhered to relevant health guidelines and education standards.

The game has 30 progressive levels (Figure 1), each incorporating five distinct challenges. In order to advance to the next level, the player must successfully complete all five challenges without making any mistakes. If a player fails to complete the challenges, the level must be replayed, reinforcing learning by ensuring that players fully understand the health content before proceeding to new topics.

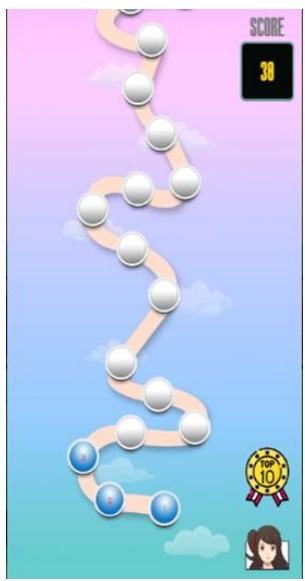

**Figure 1.** The main screen displays the 30 game levels.

The game includes a variety of interactive gameplay formats designed to engage students and facilitate learning. These formats include "*Fact or Bluff,*" "*Multiple Choice,*" "*Fill in the Blank,*" "*Word Puzzle,*" and *"Guess the Word"* (Figure 2). Additionally, players receive immediate feedback after each challenge, explaining why an answer is correct or incorrect (Figure 3). This feedback mechanism promotes active learning by reinforcing health concepts in real-time.

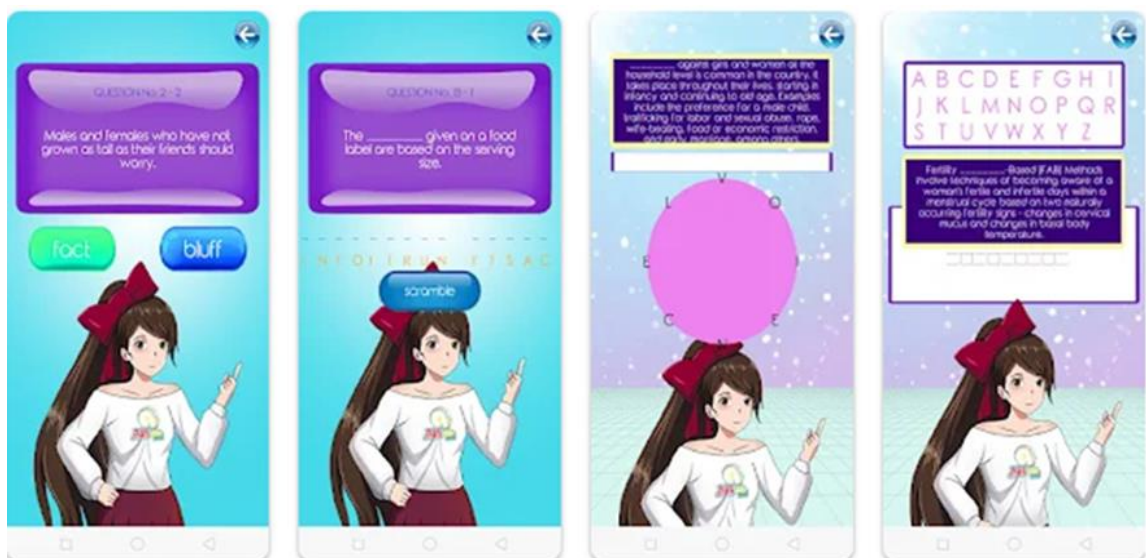

**Figure 2.** Examples of different gameplay styles within the levels.

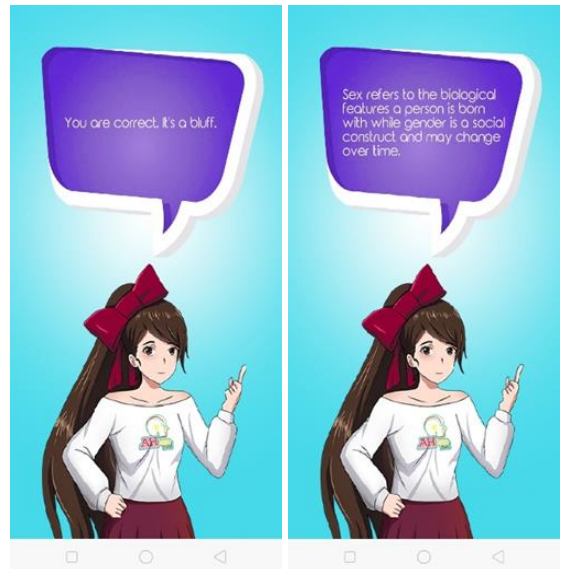

**Figure 3.** Explanation of answers provided after completing each challenge.

To further motivate players, an avatar system and item rewards were incorporated into the game. Upon successfully clearing a level, players receive random items such as shirts, jackets, pants, shoes, and accessories, which can be used to customize their avatars (Figure 4). This reward system adds an element of fun and personalization, encouraging continued gameplay and reinforcing the educational content through engagement.

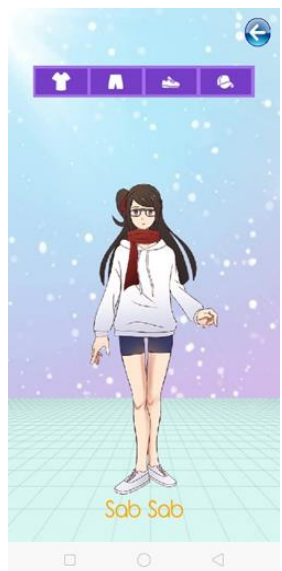

**Figure 4.** The player avatar is equipped with various customizable items earned through gameplay.

*Adolescent Health Assessment*

The primary goal of the pretest and posttest was to evaluate the effectiveness of the *AHlam Na 2.0* game application in enhancing adolescents' knowledge of health-related topics. The questions included in both the pretest and posttest were directly sourced from various levels of the game, ensuring consistency between the in-game content and the assessments.

To guarantee the reliability and accuracy of the questions, they were formulated with the expertise of health professionals from the Department of Health and the Commission on Population and Development in Region 6. These experts ensured that the questions covered critical aspects of adolescent health, including sexual and reproductive health, mental well-being, and healthy lifestyle choices.

By using the same set of questions in both the pretest and posttest, albeit in a different order, the assessment was able to measure the participants' knowledge improvement and the game's effectiveness in delivering educational content. The results from these tests serve as a key indicator of how well the game facilitated learning and whether it achieved its objective of raising health awareness among adolescents.

*Learning Satisfaction Scale*

To assess the students' acceptance and satisfaction with the *AHlam Na 2.0* mobile game, the researcher developed a learning satisfaction scale rooted in the Technology Acceptance Model (TAM). This scale evaluates four key constructs influencing technology adoption: perceived ease of use, perceived usefulness, attitude toward use, and intention to use. The questionnaire consists of 15 items, divided as follows:

a) Perceived ease of use: 4 items
b) Perceived usefulness: 4 items
c) Attitude toward use: 4 items
d) Intention to use: 3 items

Each item is rated on a five-point Likert scale, with responses ranging from strongly agree (5) to strongly disagree (1). The scale is designed to gauge the participants' learning satisfaction based on how intuitive, effective, and engaging they found the game. To ensure the reliability of the instrument, Cronbach's alpha was calculated for each construct, demonstrating strong internal consistency:

a) Perceived ease of use: 0.83
b) Perceived usefulness: 0.95
c) Attitude toward use: 0.93
d) Intention to use: 0.95

These values indicate that the learning satisfaction scale provides a reliable measure of students' acceptance and satisfaction with the game. The scale helps understand the factors contributing to the game's success in delivering health education to adolescents and offers insights into potential areas for improvement.

**C. Results and Discussion**

The study utilized Python packages such as statistics and *pingouin* for data analysis, including computations for mean, standard deviation, and t-tests (Vallat, 2018). Participants were divided into two groups based on their progress in the game: those who reached level 10 or above and those who reached below level 10. A dependent sample t-test was conducted for each group to assess knowledge improvement following the gameplay intervention.

The results for participants who reached level 9 and below demonstrate a significant improvement in their post-test scores (M = 35.00, SD = 6.91) compared to their pre-test scores (M = 32.31, SD = 6.07), suggesting that even minimal engagement with the game led to an increase in adolescent health knowledge. This finding is crucial because it indicates that the game's educational value extends beyond intensive use, meaning that students who may not be highly motivated or deeply invested in progressing through the game still manage to learn key concepts.

In terms of educational intervention, this suggests that the *"AHlam Na 2.0"* app is effective for those who are more engaged and casual users, implying a lower threshold for knowledge acquisition. This could have important implications for real-world applications, where not all students may be equally inclined to use a gamified tool intensively. The app's design seems to facilitate learning even for those who interact with it on a more superficial level, which speaks to its potential as an accessible learning tool for diverse student populations with varying levels of motivation or gaming interest. This broader reach may contribute to the widespread dissemination of adolescent health information, making the game valuable in contexts where user engagement might vary widely.

**Table 2**. Results of paired sample t-test of participants below level 10

|            | Mean  | Standard Deviation | t     |
|------------|-------|--------------------|-------|
| Pre-test   | 32.31 | 6.07               |       |
| Post-test  | 35.00 | 6.91               | -3.26 |

p<0.01

The data for participants who reached level 10 and above demonstrates a more substantial increase in knowledge than those who played the game less intensively. Specifically, their post-test scores (M = 45.45, SD = 3.59) were significantly higher than their pre-test scores (M = 35.18, SD = 4.17), as indicated by the paired sample t-test results. The large difference in the mean scores suggests that more engaged gameplay, where participants advance beyond level 10, results in a deeper understanding and retention of adolescent health information.

This substantial growth in knowledge highlights a clear correlation between the intensity of gameplay and the educational benefits gained. Participants who played the game more thoroughly, reaching higher levels, had more exposure to the health content, which likely reinforced learning and improved their comprehension of the subject matter. This suggests that the game's design, which encourages progression through levels, effectively boosts learning outcomes, particularly for users who are more engaged.

Thus, the findings indicate that the more participants interact with and engage in the game's challenges, the more pronounced their knowledge improvement becomes. This insight underscores the potential of the game to be an effective educational tool, especially when students are motivated to fully immerse themselves in the gameplay.

**Table 3**. Results of paired sample t-test of participants level 10 and above

|  | **Mean** | **Standard Deviation** | **t** |
|---|---|---|---|
| Pre-test | 35.18 | 4.17 |  |
| Post-test | 45.45 | 3.59 | -9.37 |

$p<0.01$

Table 4 reveals participants' overall positive feedback across all constructs of the learning satisfaction scale, providing key insights into the game's reception. For *perceived ease of use*, participants rated the game with a mean score of 3.73 (SD = 0.73), indicating that they found the app intuitive and easy to navigate. This suggests that the game's interface and controls were user-friendly, making it accessible to many students.

Regarding *perceived usefulness*, the app was rated 4.08 (SD = 0.86), showing that participants acknowledged the game as a valuable tool for delivering essential adolescent health information. This high score indicates that the educational objectives of the app are being met, with users recognizing its effectiveness in enhancing their health knowledge.

The construct *attitude toward use* scored a mean of 3.94 (SD = 0.82), suggesting that the game is seen as engaging and enjoyable. This result implies that students were learning and finding the process enjoyable, which is a key factor in sustaining their interest in health-related content.

Finally, *intention to use* received a lower, but still positive, mean score of 3.54 (SD = 0.88), which indicates that while most participants intend to continue using the app, there is some variation in how strongly students are motivated to keep playing the game. This might point to differences in personal interest in the topic or gaming in general.

The learning satisfaction data reveals additional insights beyond just positive feedback. The variability in the *perceived ease of use* (SD = 0.73) suggests that while many participants found the app user-friendly, some may have encountered usability challenges. Similarly, the standard deviation in *perceived usefulness* (SD = 0.86) highlights that not all participants equally regarded the app as a helpful tool for learning adolescent health, pointing to the possibility that improvements could be made to better align the game with diverse learning preferences.

The disparity between *attitude toward use* (M = 3.94, SD = 0.82) and *intention to use* (M = 3.54, SD = 0.88) is also worth noting. While participants generally had a favorable attitude toward the game, their slightly lower scores in terms of intention suggest some hesitation in continuing to use the app over time. This gap could imply a need for the app to incorporate features that encourage sustained engagement, such as more game levels, rewards, or incentives to keep users interested beyond initial play.

In sum, although the app is perceived positively, the data suggests areas for further enhancement, particularly in maintaining long-term user engagement and addressing varying user experiences with ease of use and usefulness.

Overall, the data demonstrates that the game is perceived positively in terms of ease of use, usefulness, and engagement, with a reasonable likelihood that students will continue using it in the future. These insights provide support for further development and promotion of the app as an educational tool.

Table 4. Mean and standard deviation of scores in the learning satisfaction scale

|  | Mean | Standard Deviation |
|---|---|---|
| Perceived ease of use | 3.73 | 0.73 |
| Perceived usefulness | 4.08 | 0.86 |
| Attitude toward use | 3.94 | 0.82 |
| Intention to use | 3.54 | 0.88 |

The study aims to determine the effectiveness of the "AHlam Na 2.0" application in enhancing young people's knowledge about adolescent health. The pretest-posttest results show a significant increase in students' adolescent health knowledge. It also demonstrates that the more engaged the student is in the game, the higher the adolescent health knowledge acquired through intensive gameplay. The results prove the effectiveness of the mobile game application in delivering adolescent health information to young people, showing that playing the game could raise the level of knowledge in this subject. These results correspond to the use of game-based learning in previous studies (Haruna et al., 2018; Anyi, 2019; Rincón et al., 2022; Vega et al., 2022) and demonstrate the fulfillment of the objective of the game, which is to raise awareness of adolescents in the subject matter.

Another objective of the study was to assess the application's perceived usefulness from the perspective of young people. The results indicated that most participants acknowledged the app's value and the enjoyment it brings to their learning experience. However, there was a divergence in opinions regarding the duration of gameplay, as reflected in their intentions to continue using the app.

Throughout the study, the researcher noted varied engagement levels among participants. While some students approached the game with enthusiasm and commitment, others displayed minimal interest in progressing through its levels. This observation aligns with the understanding that not all students share an affinity for gaming (Wakil et al., 2017). Future research could explore the underlying reasons for these differing behaviors toward gameplay. Additionally, although the game's design and features were developed through a design thinking process, further analysis of the application's functionality could benefit from feedback from a larger and more diverse sample population.

Lastly, this research demonstrates that the application effectively promotes adolescent health education. Therefore, it is recommended that the app be actively promoted in various regions across the Philippines to reach a wider audience and enhance health awareness among young people. Expanding its availability could significantly contribute to improving adolescent health knowledge on a national scale.

**D. Conclusion**

The findings of this study highlight the effectiveness of the "AHlam Na 2.0" mobile game in promoting adolescent health awareness among high school students. The research demonstrated significant improvements in participants' knowledge of health-related topics after engaging with the mobile game. The results indicated that both casual and intensive gameplay positively impacted the participants' knowledge acquisition, suggesting that the game can cater to students with varying levels of engagement and motivation. The analysis revealed that students who played to an advanced level showed a more substantial improvement in their post-test scores compared to those who played less intensively. This underscores the importance of continuous interaction with the game's educational content for deeper learning and retention. The learning satisfaction data further supports the game's effectiveness as an educational tool. The high scores in perceived usefulness and attitude toward use indicate that students recognized the game's value and enjoyed the learning process. However, the slightly lower score in intention to use suggests that additional features, such as more engaging rewards or continued updates, could help sustain long-term interest and engagement with the app.

Overall, the findings of this study confirm that the "AHlam Na 2.0" mobile game is an effective and accessible educational tool for promoting adolescent health awareness. The study suggests that integrating gamified learning interventions into health education curricula can significantly enhance students' knowledge retention and overall learning satisfaction. Future improvements to the app, particularly in sustaining long-term user engagement, can further optimize its impact as a digital learning resource for adolescents.

### E. Acknowledgment

I extend my heartfelt gratitude to the Commission on Population and Development Regional Office VI, Iloilo, Philippines for giving me the invaluable opportunity to be part of the AHlam Na 2.0 development team. Your support and collaboration played a crucial role in the success of this initiative.

I am deeply thankful to the Iloilo State University of Fisheries Science and Technology (formerly Iloilo State College of Fisheries), especially to our team leader Prof. Mary Sol R. Baldevarona, for the trust and confidence placed in me throughout the project. Your unwavering support greatly contributed to the completion of this research endeavor.

A special mention of appreciation goes to Dingle National High School, Dingle, Iloilo, Philippines, for graciously allowing me to conduct this activity. Your cooperation and participation have made a significant impact on the realization of this paper, and for that, I am truly grateful.

This accomplishment would not have been possible without the collective efforts of these institutions, whose contributions have been invaluable to the project's success and its potential to make a positive difference in adolescent health education.

Caliston, N.